\documentclass[11pt]{article}
\usepackage{moriond,epsfig}

\def\Journal#1#2#3#4{{#1} {\bf #2}, #3 (#4)}

\def\PRL{\em Phys. Rev. Lett.}
\def\PRD{{\em Phys. Rev.} D}

\begin{document}
\vspace*{4cm}
\title{$B$ PHYSICS RESULTS FROM CLEO}
\author{M. R. SHEPHERD \\ (for the CLEO Collaboration)}
\address{Laboratory for Elementary-Particle Physics, Cornell University, Ithaca, NY, 14853 USA}
\maketitle

\abstracts{ 
We present results on lepton energy and recoil hadronic mass moments in semileptonic $B$ decay using a total of 9.4 fb$^{-1}$ of data taken with the CLEO detector at the $\Upsilon\left(4S\right)$.  These results are discussed in the context of Heavy Quark Effective Theory and compared to theory predictions as a function of the minimum lepton energy requirement.  We also measure the $B$ semileptonic branching fraction, $\mathcal{B}\left(B\to Xe^+\nu_e\right)$, as $\left(10.91\pm0.09\pm0.24\right)\%$.
 }

\section{Inclusive Semileptonic $B$ Decay}

Experimentalists have long used semileptonic $B$ decay as a place to measure the magnitude of the elements $V_{ub}$ and $V_{cb}$ of the CKM quark-mixing matrix.   In the case of inclusive semileptonic measurements, where only
the lepton and neutrino are studied, a significant source of error in the measurements of CKM matrix elements arises from the lack of understanding of the dynamics of the $b$ quark inside the $B$ meson.  The non-perturbative interaction of the $b$ quark with the light degrees of freedom of the meson leaves its imprint on the spectra of kinematical variables measured in inclusive semileptonic decay.

Heavy Quark Effective Theory (HQET) combined with QCD implemented in an Operator Product Expansion (OPE) has provided a framework to understand how the $b$ quark interacts with the light degrees of freedom of the meson.  The expansion combines QCD with HQET to handle short and long range interactions within the meson.  The result is that inclusive observables can be expressed in powers of $m_B^{-1}$, where $m_B$ is $B$ meson mass.  This expansion has one free parameter at $\mathcal{O}\left(m_B^{-1}\right)$, $\bar{\Lambda}$, where
\begin{equation}
\bar{\Lambda} = m_B - m_b|_{m_b\to\infty}
\end{equation}
and represents the energy of the light cloud.  At $\mathcal{O}\left(m_B^{-2}\right)$, two more parameters, $\lambda_1$ and $\lambda_2$ appear.  The parameter $\lambda_1$ is related to the average momentum squared of the $b$ quark in the rest frame of the meson, and $\lambda_2$ describes the spin-dependent chromo-magnetic interaction of the $b$ quark with the light cloud which can be precisely determined from $B/B^*$ and $D/D^*$ mass splittings.    We should note that these two parameters are related to the frequently used parameters $\mu_\pi^2$ and $\mu_g^2$ by $\lambda_1\equiv -\mu_\pi^2$ and $\lambda_2\equiv\mu_g^2/3$. 

Expressions for the moments of kinematical variables in inclusive semileptonic $B$ decay at $\mathcal{O}\left(m_B^{-2}\right)$ explicitly contain these parameters.  Therefore, measuring many such moments constrains the parameters and provides a mechanism to validate the theory.

\section{Recoil Hadronic Mass Moments}

\begin{figure}
\begin{center}
\epsfig{file=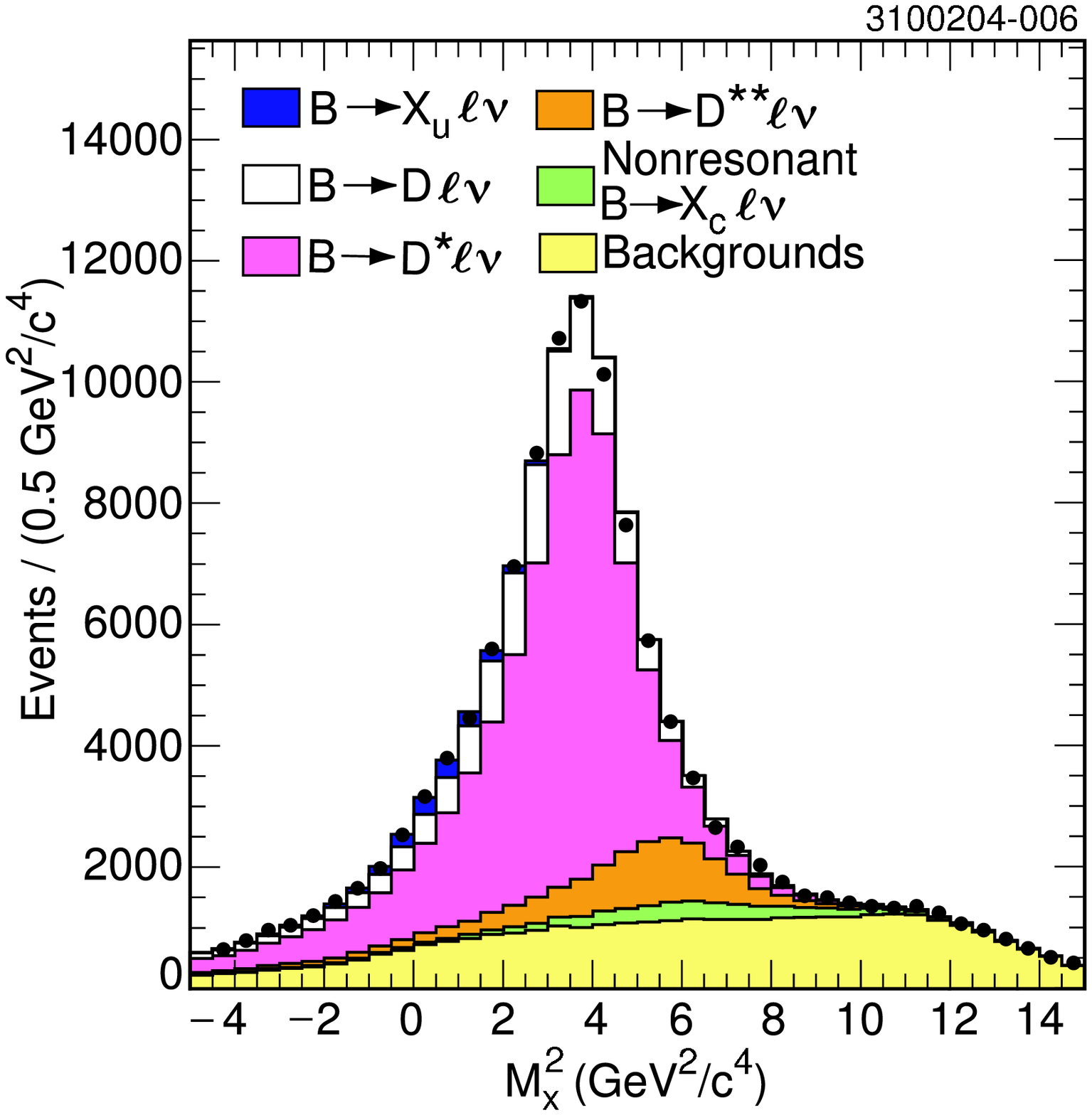, scale=0.4}\epsfig{file=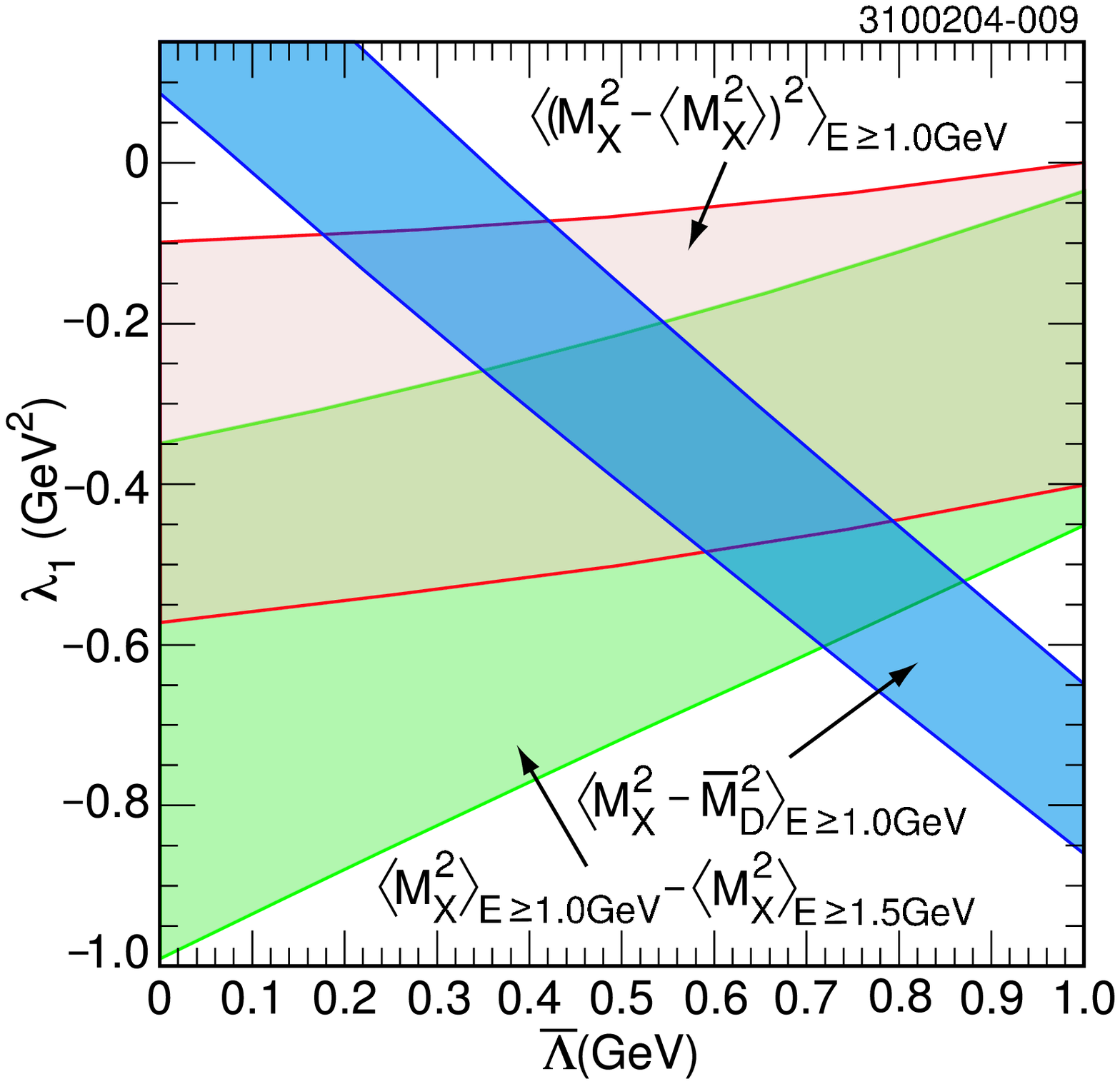,scale=0.43}
\end{center}
\caption{ (Left) The $M_X^2$ projection of the of the three-dimensional fit to the decay spectrum.  Fit components are shown in the histograms while the data are represented by the points.  (Right) Constraints on the HQET parameters, $\bar{\Lambda}$ and $\lambda_1$, with approximate $1\sigma$ combined theory and experimental errors. }
\label{fig:mx2}
\end{figure}

The hermetic CLEO detector and clean initial state allows one to ``recontruct" neutrinos from the missing energy and momentum.  By reconstructing the lepton and neutrino from a semileptonic $B$ decay we measure the full triply-differential decay rate\cite{elliot} in terms of the kinematical variables $q^2$, $M_X^2$, and $\cos \theta_{Wl}$, which are the mass squared of the virtual $W$, the mass squared of the recoil hadronic system, and the the angle between the lepton in $W$ rest frame and $W$ in the $B$ rest frame (assumed to be the lab frame).

A three dimensional fit is performed to the measured spectrum.  Components of the fit consist of Monte Carlo modeled signal $B$ semileptonic decay modes along with estimated background contributions from sources such as fake or secondary leptons.  The $M_X^2$ projection of the fit is shown in Figure~\ref{fig:mx2} and the agreement with data is excellent.

The analysis is repeated with a variety of values for the minimum lepton energy.  Below we shows results for the first and second moments of the $M_X^2$ distribution for lepton energy cuts of 1.0 and 1.5 GeV.  The errors are statistical, experimental systematic, and model dependence uncertainties.  In Figure~\ref{fig:mx2} we also show the constraints placed on the HQET parameters $\lambda_1$ and $\bar{\Lambda}$ using the pole mass scheme of Bauer {\it et. al} \cite{bauer}.

\begin{center}
\begin{tabular}{|c|c|c|}\hline
Moment & $E_l>1.0$ GeV & $E_l>1.5$ GeV \\\hline
$\left\langle M_X^2 - \bar{M}_D^2\right\rangle$ (GeV$^2$/$c^4$) & $0.456\pm0.014\pm0.045\pm0.109$ & $0.293\pm0.012\pm0.033\pm0.048$ \\
$\left\langle\left(M_X^2 - \left\langle M_X^2 \right\rangle\right)^2\right\rangle$ (GeV$^4$/$c^8$) & $1.266\pm0.065\pm0.222\pm0.631$ & $0.629\pm0.031\pm0.088\pm0.113$ \\\hline
\end{tabular}
\end{center}

\section{Lepton Energy Moments}

\begin{figure}
\begin{center}
\epsfig{file=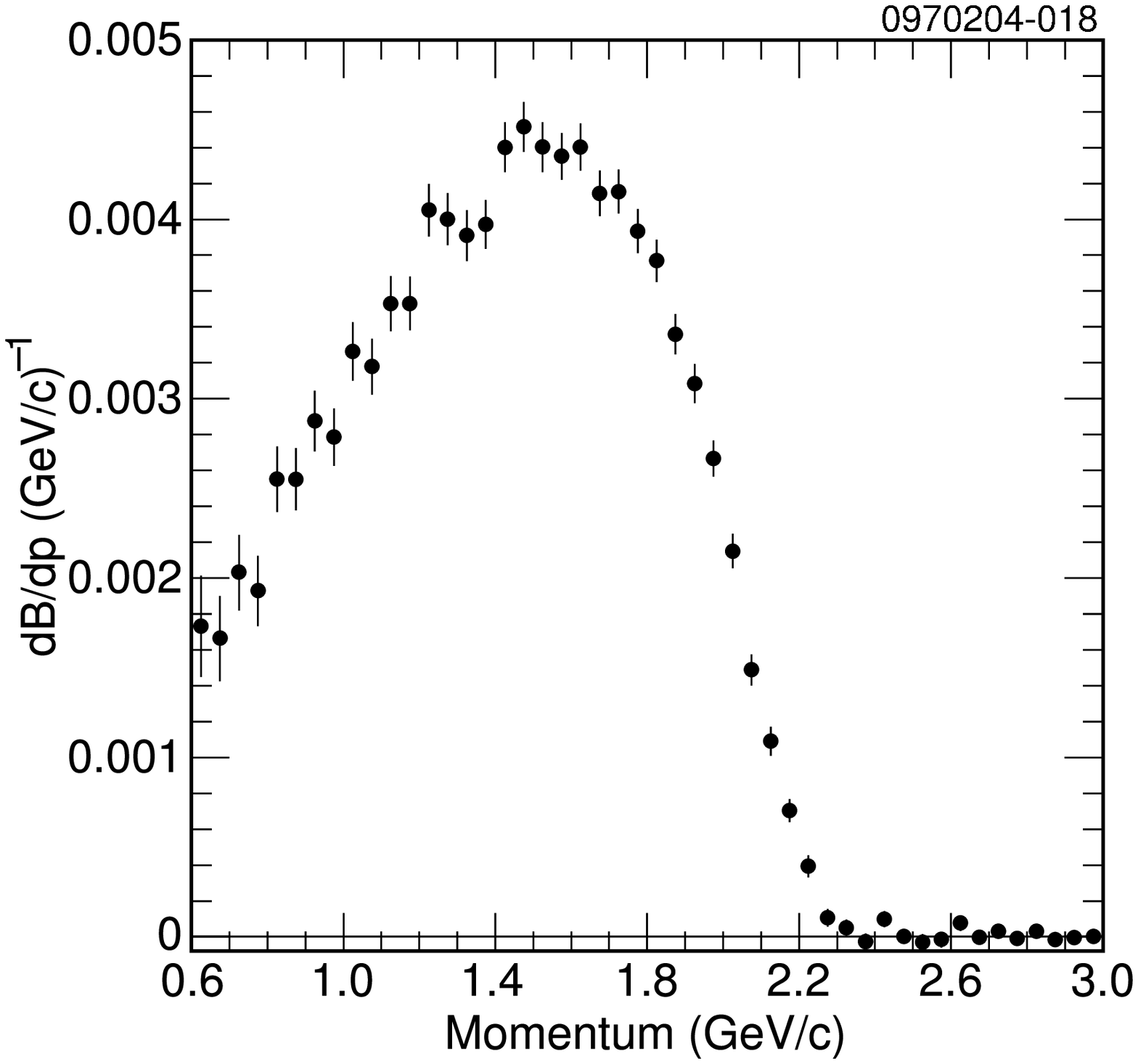, scale=0.42}\epsfig{file=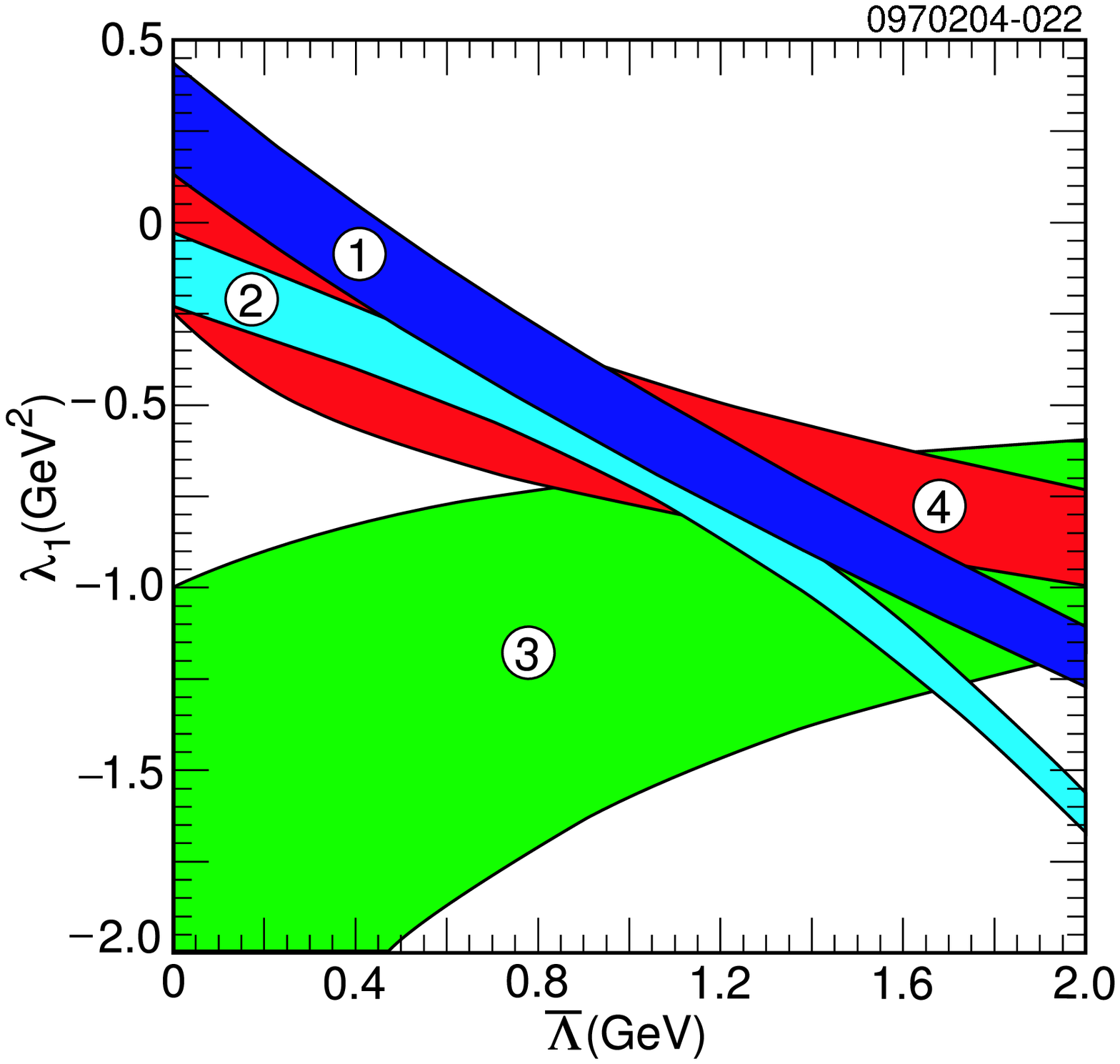,scale=0.42}
\end{center}
\caption{ (Left) The primary electron energy spectrum.  (Right) Constraints on the HQET parameters, $\bar{\Lambda}$ and $\lambda_1$, with approximate $1\sigma$ combined theory and experimental errors determined by 
$\left\langle E_l\right\rangle_{0.7}$ (1), 
$\left\langle E_l^2 - \left\langle E_l\right\rangle^2\right\rangle_{0.7}$ (2),
$\left\langle E_l\right\rangle_{1.5} - \left\langle E_l\right\rangle_{0.7}$ (3), and
$\left\langle E_l^2 - \left\langle E_l\right\rangle^2\right\rangle_{1.5} - \left\langle E_l^2 - \left\langle E_l\right\rangle^2\right\rangle_{0.7}$ (4).}
\label{fig:el}
\end{figure}

In general there are two principal sources of leptons in $B$ decay:  those originating from $b\to c l \nu$ processes, ``primary" leptons, and those originating from $b\to c\to s l \nu$ processes, ``secondary" leptons.  Secondary leptons are typically much lower momentum than primary leptons.  By measuring the primary electron spectrum and its moments we can determine the overall $B\to Xe^+\nu_e$ branching fraction and place constraints on the HQET parameters mentioned above.  

The sign of a primary lepton uniquely identifies the parent quark as a $b$ or $\bar{b}$ quark.  A typical $\Upsilon(4S)$ decay produces two $B$ mesons.  We search for a high-momentum primary lepton from the decay of one $B$ and use it to ``tag" the flavor of the meson.  We then look for additional electrons in the event coming from semileptonic decay of the other $B$ and compare their charge with that of the tag lepton.  These additional electrons are classified as ``like-sign" or ``unlike-sign."  In unmixed events primary electrons from the other $B$ are unlike-sign while secondary electrons are like-sign.  In events where one of the neutral $B$ mesons mixes, the opposite is true.  A significant background in the unlike-sign spectrum comes from taking both the primary and secondary lepton from the same $B$.  In this case the leptons are kinematically correlated and this background can be successfully eliminated with a cut on the angle between the leptons.

We measure both the like-sign and unlike-sign signal electron spectra and can represent these two spectra in terms of the primary and secondary spectra as:
\begin{eqnarray}
\label{eq:ulsign}
\frac{dN\left(l^\pm e^\mp\right)}{dp} &=& N_l\eta(p)\epsilon(p)\left[\left(1-\chi\right)\frac{d\mathcal{B}(b\to e)}{dp} + \chi\frac{d\mathcal{B}(b\to c\to e)}{dp}\right] \\
\label{eq:lksign}
\frac{dN\left(l^\pm e^\pm\right)}{dp} &=& N_l\eta(p)\left[\chi\frac{d\mathcal{B}(b\to e)}{dp} + \left(1-\chi\right)\frac{d\mathcal{B}(b\to c\to e)}{dp}\right] 
\end{eqnarray}
where $\eta(p)$ and $\epsilon(p)$ are the efficiencies for the signal electron and the cut to remove same-$B$ secondary leptons.  $N_l$ is the number of tag leptons and $\chi$ is the $B$ mixing parameter times the fraction of $\Upsilon(4S)$ events that produce neutral $B$ mesons.  While not depicted in the equations above, we also allow for the possibility that the secondary spectra differ in charged and neutral $B$ decays.

Using equations~\ref{eq:ulsign} and~\ref{eq:lksign} and the measured like- and unlike-sign spectra we can extract the primary electron spectrum for semileptonic $B$ decay\cite{chris}, shown in Figure~\ref{fig:el}.  The zeroth moment of the primary spectrum gives a measurement of $\mathcal{B}\left(B\to Xe^+\nu_e\right)$, as $\left(10.91\pm0.09\pm0.24\right)\%$.  The first and second moments of the spectrum can be computed with various lepton energy momentum cuts and are shown below.  In Figure~\ref{fig:el} we also show the constraints placed on the HQET parameters $\lambda_1$ and $\bar{\Lambda}$ by these moments using the pole mass scheme of Bauer {\it et. al} \cite{bauer}.

\begin{center}
\begin{tabular}{|c|c|c|}\hline
Moment & $E_l>0.7$ GeV & $E_l>1.5$ GeV \\\hline
$\left\langle E_l\right\rangle$ & $1.4509\pm0.0035\pm0.0079$ & $1.7792\pm0.0021\pm0.0027$ \\
$\left\langle E_l^2 - \left\langle E_l \right\rangle^2\right\rangle$ & $0.1374\pm0.0015\pm0.0018$ & $0.0316\pm0.0008\pm0.0010$ \\\hline
\end{tabular}
\end{center}

\section{Comparison With HQET Predictions}

\begin{figure}
\begin{center}
\epsfig{file=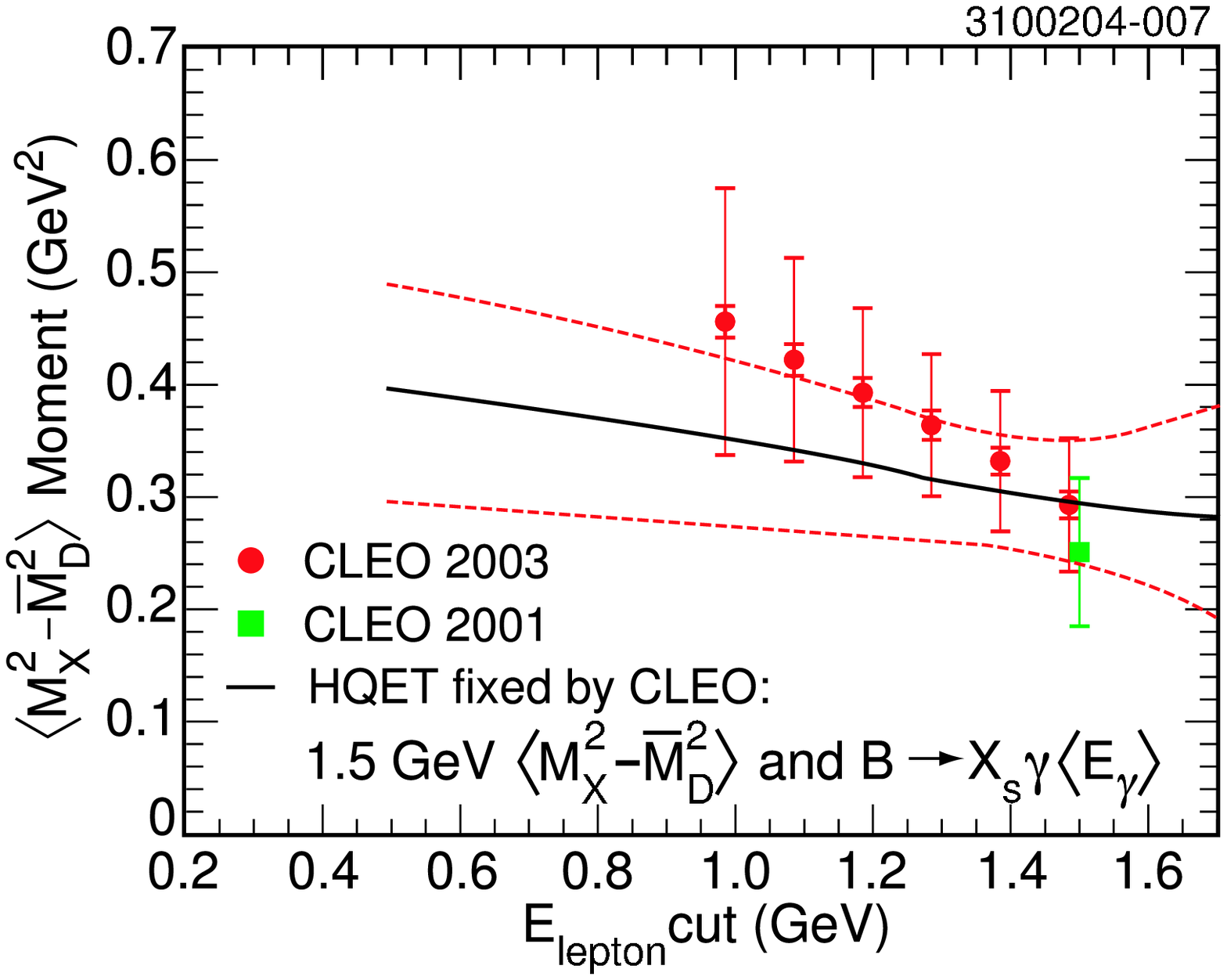, scale=0.45}\epsfig{file=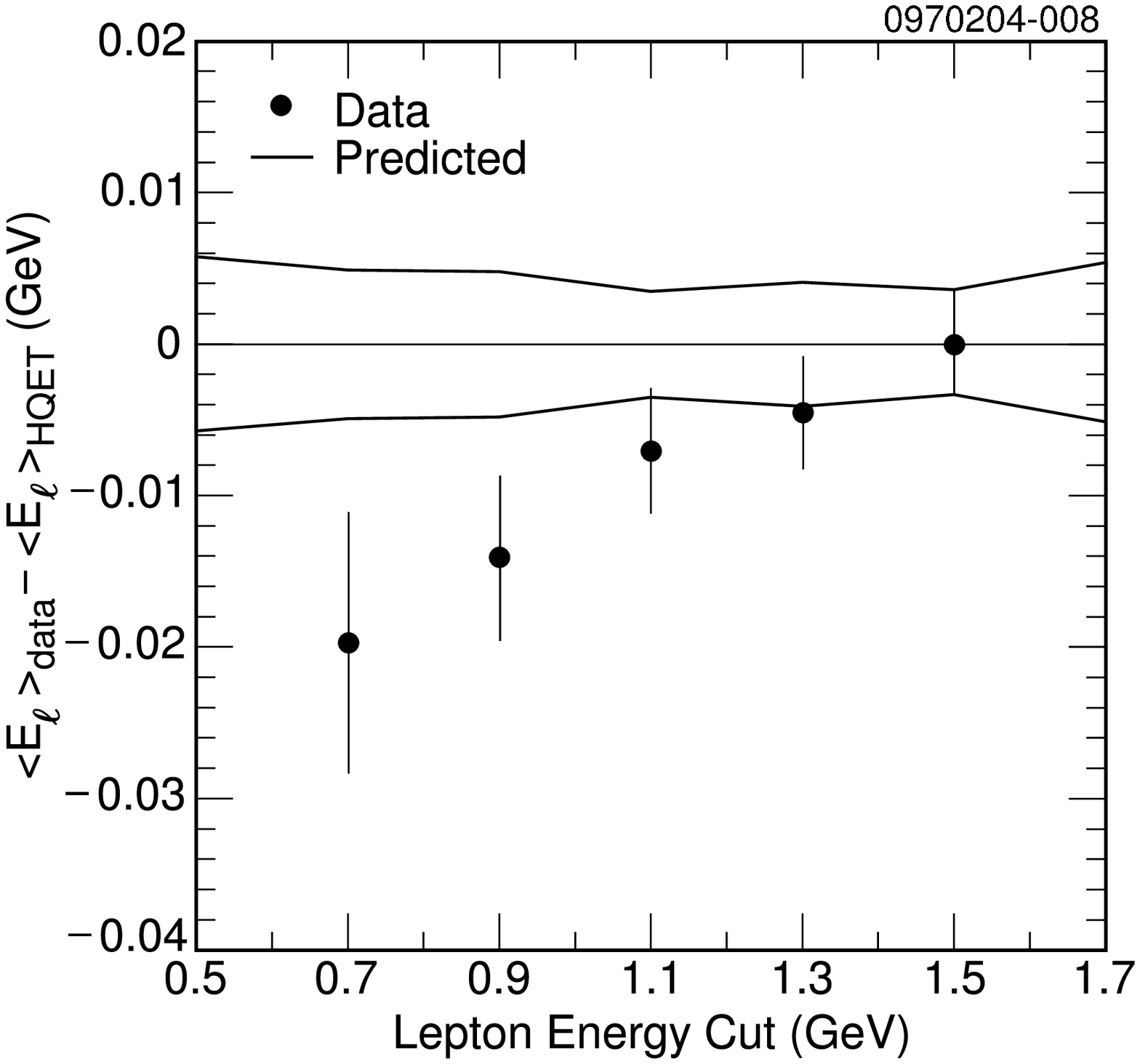,scale=0.37}
\end{center}
\caption{ Comparison of measured first moments of the $M_X^2$ (left) and $E_l$ (right) distributions as a function of minimum lepton energy cut with theory predictions.  In each case theory is fixed using 1.5 GeV point and the CLEO $b\to s\gamma$ measurement. }
\label{fig:vcut}
\end{figure}

We can use the data with a 1.5 GeV minimum lepton energy cut along with the CLEO $b\to s\gamma$ measurement\cite{bsg} to constrain the values of the HQET parameters.  Using these values in the Bauer pole mass scheme \cite{bauer}, we can predict the moments at lower lepton energy cut.  Figure~\ref{fig:vcut} shows this prediction along with the measured moments at various lepton energy cuts.  A discrepancy emerges in the first moment of the $E_l$ spectrum as the cut is lowered.

\section*{Acknowledgments}
We would like to thank the CESR staff for providing us with luminosity to make this measurement and the National Science Foundation for their financial support.  I would like to thank my CLEO colleagues, especially Chris Stepaniak, Elliot Lipeles, and Dan Cronin-Hennessy for their productive discussions.

\end{document}